\documentclass[twocolumn,showpacs,aps,preprintnumbers,amsmath,amssymb,superscriptaddress]{revtex4}
\usepackage{graphicx}
\usepackage{dcolumn}
\usepackage{bm}
\begin{document}
\preprint{ Physical Review Letters 102, 045901, 2009}
\title{Anomalous diffusion mediated by atom deposition into a porous substrate.}
\author{Pascal Brault}
\email[corresponding author:]{Pascal.Brault@univ-orleans.fr}
\affiliation{Groupe de Recherches sur l'Energ\'etique des Milieux Ionis\'es,
UMR6606 CNRS-Universit\'e d'Orl\'eans BP 6744, 45067 Orl\'eans Cedex 2, France}
\author{Christophe Josserand}
\affiliation{Institut Jean Le Rond d'Alembert, UMR 7190 CNRS-Paris VI Case 162, UPMC, 4 place Jussieu, 75252 Paris 
Cedex 05,  France}
\affiliation{Kavli Institute for Theoretical Physics, University of California, Santa Barbara, CA 93106,USA}
\author{Jean-Marc Bauchire}
\affiliation{Groupe de Recherches sur l'Energ\'etique des Milieux Ionis\'es,
UMR6606 CNRS-Universit\'e d'Orl\'eans BP 6744, 45067 Orl\'eans Cedex 2, France}
\author{Ama\"{e}l Caillard}
\affiliation{Groupe de Recherches sur l'Energ\'etique des Milieux Ionis\'es,
UMR6606 CNRS-Universit\'e d'Orl\'eans BP 6744, 45067 Orl\'eans Cedex 2, France}
\affiliation{Space Plasma, Power, and Propulsion group (SP3), Research School of Physical Sciences and Engineering, The 
Australian National University, Canberra, ACT 0200, Australia}
\author{Christine Charles}
\affiliation{Space Plasma, Power, and Propulsion group (SP3), Research School of Physical Sciences and Engineering, The 
Australian National University, Canberra, ACT 0200, Australia}
\author{Rod W. Boswell}
\affiliation{Space Plasma, Power, and Propulsion group (SP3), Research School of Physical Sciences and Engineering, The 
Australian National University, Canberra, ACT 0200, Australia} 
\begin{abstract}
Constant flux atom deposition into a porous medium is shown to generate a dense overlayer and a diffusion profile. 
Scaling analysis shows that the overlayer acts as a dynamic control for atomic diffusion in the porous substrate. This 
is modeled by generalizing the porous diffusion equation with a time-dependent diffusion coefficient equivalent to a 
nonlinear rescaling  of time.
\end{abstract}
\pacs{66.30.Pa, 66.30.Dn, 68.55.-a, 81.15.Cd}

\maketitle
Thin film depositions on substrates are important in many physical processes and applications. Moreover, deposition on porous substrates is particularly useful for catalytic systems \cite
{bra1,cail} in which atomic deposition carried out by plasma sputtering is then coupled with 
the diffusion of the atoms into the porous substrate. 
Diffusion in a porous medium is generally anomalous and is characterized by the mean square 
displacement $<z^2(t)>$: it evolves with a power law in time $t^\alpha$ different from the well known linear behavior (
$\alpha=1$) for normal diffusion.
Anomalous diffusion \cite{bou,metz} is a general process that can be observed in many domains such  as transport in 
porous \cite{aron} and/or fractal media \cite{step}, surface growth \cite{spoh}, solid surface diffusion \cite{LuLa99} 
or hydrodynamics (rotating flows \cite{WeSw98}, turbulence \cite{Ric26} or diffusion in an array of convection rolls 
\cite{CaTa88,YoPu89}). It can be described by different models that involve space-dependent diffusion 
coefficients \cite{bou,metz,mala,pedr,tsal,shau1,shau2}. In these models anomalous diffusion is included through the propagator 
(the solution of the diffusion equation starting with a Dirac distribution at $t=0$) that exhibits both power laws and 
stretched exponential behaviors.
Here, we study platinum deposition by plasma sputtering on a porous carbon substrate. The platinum atoms are sputtered 
by the plasma ions and travel to the substrate so that both deposition and diffusion processes can depend on plasma operating parameters.
The experimental results consist in a measure of the density profiles  of the deposited matter at different times. 
The goal of this Letter is to characterize this diffusion-deposition process in a porous carbon medium using the 
time evolution of the platinum density profile.

\begin{figure}
\includegraphics[scale=0.3]{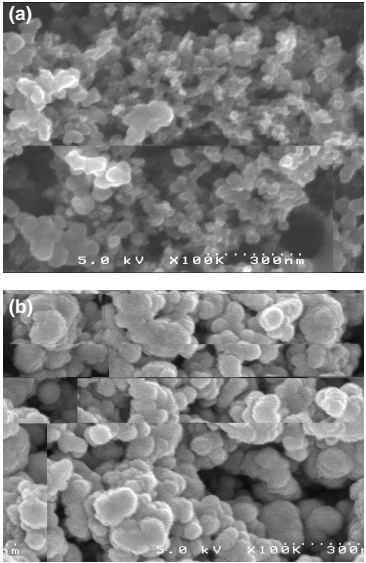}
\caption{Scanning Electron Microscopy top view of porous carbon layer a) before Pt deposition b) after 
Pt deposition (\em Courtesy D. Cot, IEMM CNRS-ENSCM-Univ. Montpellier II)}
\label{fig.1}
\end{figure}

\begin{figure}
\includegraphics[scale=0.3,angle=-90]{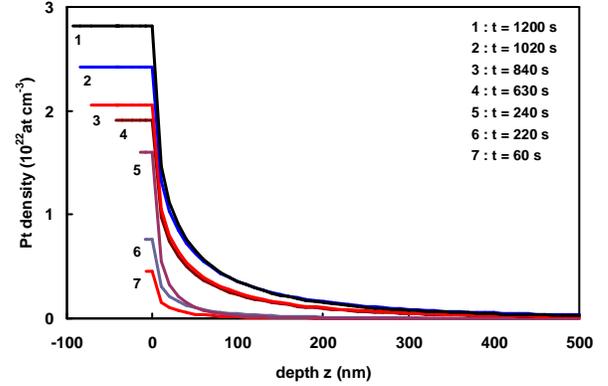}
\caption{(Color online) Pt depth profiles in porous carbon at different successive deposition times. $z=0$
corresponds to the porous carbon layer surface}
\label{fig.2}
\end{figure}

\begin{figure} 
\includegraphics[scale=0.4]{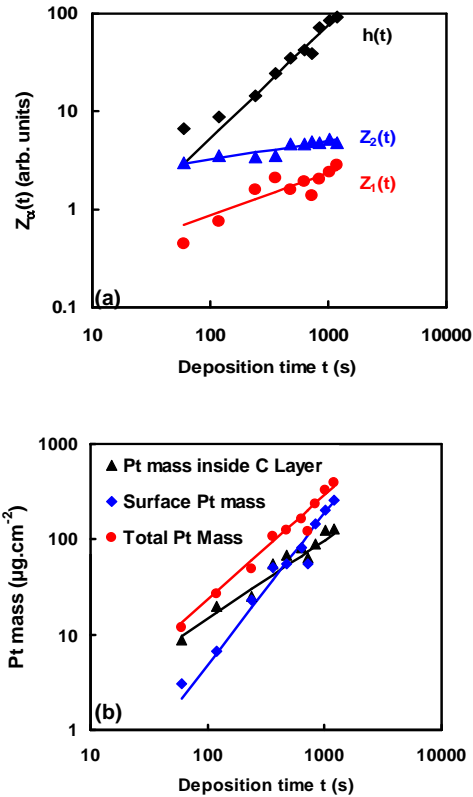}
\caption{(Color online) For the $0.5$ Pa deposition, a) thickness of the Pt layer $h(t)$ above the porous medium and 
value of $Z_1(t)$ 
and $Z_2(t)$ from the fitted law (\ref{fit-exp}) as a function of time (log-log plot). The lines correspond to the 
power laws $Z_1(t) \sim t^{0.4}$,$Z_2(t) \sim t^{0.2}$ and $h(t) \sim t^{1.1}$; b) Amount of Pt mass in the porous 
medium calculated using (\ref{fit-exp}) as a function of time (log-log plot).} 
\label{fig.3}
\end{figure}

Platinum atoms are deposited by plasma sputtering  into a porous carbon layer supported (See Fig.~
\ref{fig.1}) on a carbon cloth. This porous layer is a few tens of  microns thick and is composed of randomly stacked 
carbon nanoparticles (Vulcan XC 72) and PTFE particles brushed onto the
carbon cloth. The specific area is 15 m$^2$g$^{-1}$ before Pt deposition and slightly lower at
 13 m$^2$g$^{-1}$ after Pt deposition. Examination of Fig.\ref{fig.1} shows the pore size reduction.  Platinum atoms 
are deposited onto this porous carbon layer and diffuse in the course of deposition. 
The platinum deposition reactor is a previously described \cite{bra1,cail} plasma sputter system which delivers a
constant deposition rate. An argon plasma is created in the stainless steel deposition chamber $18$ cm inner-diameter and
$25$ cm-long by using an external planar antenna (also known as TCP antenna).
The porous layer is placed on a movable grounded substrate holder in front of the platinum sputter target with a target-substrate distance of $4.5$ cm. 
The porous  layer is thus exposed to a flux of sputtered platinum, into which it diffuses under the operating conditions. 
The diffusion process was studied at argon pressures of 0.5 and 5 Pa, leading to a mean kinetic energy 
$E_k$ of the sputtered 
Pt atom around 7 eV  \cite{cail} and 0.04 eV respectively \cite{cail2}. The resulting platinum depth profiles were
measured using Rutherford Backscattering Spectroscopy,
which gives an indirect measure of the average density as a function of the depth $z$. The density is obtained 
by fitting the experimental spectrum with a spectrum derived from a defined profile function.
Firstly, we observe that the platinum profile can in fact be decomposed in two distinct regions: a growing layer above 
the porous medium, of mean thickness $ h(t) = z_0(t)$ and constant density $\rho(z,t) = Z_1(t), -z_0(t) < z \leq 0 $ and 
a density profile $\rho(z,t), z > 0$ in the porous medium (which extends in the $z>0$ 
domain, $z=0$ corresponding to the porous interface). 
The Pt depth profile is thus deduced by minimizing the difference between the experimental RBS spectrum and the 
simulated RBS spectrum using the profile defined by Eq. \ref{fit-exp}. It corresponds to the known solution 
of anomalous diffusion processes \cite{cail} i.e. a stretched exponential. 
\begin{eqnarray}
\rho(z,t) = Z_1(t), & \qquad -z_0 < z \leq 0 \nonumber \\
\rho(z,t) = Z_1(t)e^{-\frac{z^{2+\theta}}{Z_2(t)}}, & z > 0 
\label{fit-exp}
\end{eqnarray}
where $\theta$ is the dimensionless coefficient which characterizes the anomalous diffusion behavior \cite
{mala,pedr,tsal}.This fitting procedure leads to the error bar of the exponent $\theta$ as reported hereafter.
The density profile is continuous at the interface between the growing platinum layer and the porous 
medium at $z=0$ so that we have $\rho(0,t)=Z_1(t)$. $Z_2(t)$ is the time dependent spreading of the stretched 
exponential. Such a profile suggests that the particle flux delivered by the 
plasma sputtering cannot be absorbed directly by the porous layer with the result that a fraction of the incident 
atoms have to deposit at the interface. It also implies that the diffusion process into the porous layer is a 
consequence of both the diffusion of the platinum atoms that have penetrated into the substrate and the 
diffusion of those absorbed at the interface. \\
Some fitted platinum depth profiles at different deposition times $t$ in the porous medium are displayed in Fig. \ref
{fig.2}. 
The best fits correspond to $\theta = -1.45 \pm 0.05 \approx -3/2$ which suggests a super-diffusive
behavior in the porous medium (see below). Moreover, given $\theta = -1.5$, $Z_1(t)$ and $Z_2(t)$ asymptotically follow 
power law behaviors with time ($Z_1 \sim t^m$ and $Z_2 \sim t^p$) with $m$ being $0.40\pm 0.05$ and $p$ being $0.20 \pm 
0.05$ as shown in Fig. \ref{fig.3}(a). $h(t)$ also obeys a less well
defined power law behavior with $\sim t^n, n=1.1 \pm 0.1$ . The  mass of platinum inside the porous medium is found to scale as 
$t^{0.8\pm0.1}$ while the Pt mass of the surface overlayer, $Z_1(t).h(t)$, scales as $t^{1.5\pm0.1}$ as shown on Fig. 
\ref{fig.3}(b). However, we obtain an almost linear evolution of the total mass ($\sim t^{1.1\pm0.1} $) of platinum 
deposited in and on the porous medium, as expected by the experimental process. At the higher pressure, 5 Pa,
$\theta = -1.33 \pm 0.05$,$m = p = 0.25 \pm0.05$ and $n=1.0 \pm0.1$. Hence, a change of deposition conditions yields a change of Pt concentration profile within the porous carbon.  
To help explain our results, we use a classical model of the anomalous diffusion 
process \cite{mala,pedr,tsal,shau1} in fractal and porous media, considered here in one dimensional space accounting 
for the in plane averaging (we can discard nonlinear diffusion in the porous media from the experimental results) resulting in the following generalized diffusion equation for the density profile $\rho(z,t)$ for $z\geq 0$:
\begin{equation}
\frac{\partial \rho}{\partial t}= \frac{\partial}{\partial z} \left(\frac{K_0}{z^{\theta}} \frac{\partial \rho}{\partial
z}\right)
\label{diffeq}
\end{equation}
where $\theta$ is a real parameter characterizing anomalous diffusion and is {\it a priori} unknown.
This equation is obtained by postulating a distance dependence $\frac{K_0}{z^{\theta}}$ of the diffusion 
coefficient in Fick's law. Classical 
Brownian diffusion holds for $\theta=0$, whereas positive (negative) $\theta$ corresponds to sub(super)-diffusive 
dynamics where the mean-square displacement scales as $<z^2> \propto 
t^{2/(2+\theta)}$.  The general solution for this equation (the so-called propagator) exhibits a power law and stretched 
exponential behavior and is given by \cite{metz,mala,pedr,tsal,shau1,shau2}:
\begin{equation}
\rho(z,t) \propto \left[ K_0 \left(2+\theta\right)^2 t \right]^{-1/(2+\theta)} 
\exp \left[ -\frac{z^{2+\theta}}{K_0(2+\theta)^2 t} \right]
\label{soleqdiff}
\end{equation}
which holds for a normalization condition consistent with mass conservation.
Since the deposition is carried out with a constant flux, the total mass is expected {\it a priori} to grow linearly 
and the solution to our problem should obey the boundary condition at $z=0$:
\begin{equation}
	\frac{K_0}{z^\theta} \frac{\partial \rho}{\partial z}= C_0
\label{BC}
\end{equation}
where $C_0$ is the constant flux of platinum imposed by the plasma sputtering.
Therefore, the solution can be obtained using the propagator solution (\ref{soleqdiff}) through a linear superposition 
consistent with the boundary condition (\ref{BC}). It is in fact more convenient to investigate the solution in the 
following self-similar form:
\begin{equation}
\rho(z,t)=t^{\beta} f\left(\frac{z}{t^\alpha} \right).
\label{self}
\end{equation}
While the propagator obeys $\alpha+\beta=0$ (constant mass solution), the constant flux solution must have $\alpha + 
\beta = 1$. Inserting (\ref{self}) into the diffusion equation (\ref{diffeq})
we obtain $\alpha=\frac{1}{2+\theta}$ while $\beta$ is fixed by the flux condition.
The self-similar function $f(\xi)$ is then the solution of the ordinary differential equation ($\xi=z/t^\alpha$ being 
the self-similar variable):
\begin{equation}
	\beta f-\alpha \xi f'=\frac{K_0}{\xi^{\theta+1}} \left (\xi f''-\theta f' \right)
\end{equation}
which gives the propagator solution (\ref{soleqdiff}) for $\alpha+\beta=0$. Moreover, it can be seen that the 
density profile for large $z$ follows the stretched exponential law (\ref{fit-exp}) with $Z_1(t) \propto t^\beta$ and 
$Z_2(t)=K_0(2+\theta)^2t$.\\
The predictions of this simple model do not agree with the experimental results: from $\theta 
= -1.5$, one would obtain $\alpha=2$ and $\beta=-1$ since we have the condition $\alpha+\beta=1$. Hence, there is a marked discrepancy between the experimental measurements ($\alpha\sim 0.4$ and 
$\beta \sim 0.4$) and the model even when the error bars of the experiments are taken into account. However, the 
underlying physical processes of atomic diffusion and the measured density profiles suggest that the general diffusion 
equation proposed here is a good framework for modeling atomic deposition through plasma sputtering. 
It is thus tempting to model the effect of the growing external platinum layer and of the plasma by keeping 
the same equation with the anomalous diffusion coefficient $K(t)$ and a flux of mass at $z=0$, both time dependent. 
Although we are far from a detailed microscopic derivation of the model, it is postulated here that the growth 
of the platinum layer has a screening effect on the diffusion and thus alters the mass flux towards the porous medium. 
In particular the platinum overlayer acts as a reservoir for diffusion. In addition, the temperature
distribution, not accounted for directly by this model, is certainly influenced by the presence of the
platinum layer. Since the diffusion coefficient is a function of the temperature, it is probably time dependent. This 
time dependent diffusion coefficient can also be understood as a time rescaling and it is needed in the framework of 
the diffusive equation since it is the only way to change the time dependence in the exponential  law (\ref
{soleqdiff}). We thus propose to model the deposition/diffusion dynamics through the previous equation (\ref
{diffeq}) with a time dependent diffusion coefficient $K(t)$:
\begin{equation}
\frac{\partial \rho}{\partial t}= \frac{\partial}{\partial z} \left(\frac{K(t)}{z^{\theta}} \frac{\partial \rho}{
\partial z}\right)
\label{gen-eq}
\end{equation}
The presence of the external growing layer is also accounted for by an {\it a priori} time dependent boundary condition:
\begin{equation}
	\frac{K(t)}{z^\theta} \frac{\partial \rho}{\partial z}= C(t)
\label{BCgen}
\end{equation}
To illustrate the results with no loss of generality, we will seek a power law behavior for a sufficiently time. We 
introduce two additional exponents $\epsilon$ and $\gamma$ such that the flux of atoms and the diffusion 
coefficient follow:
\begin{equation}
C(t)=C_0 t^{\epsilon} \qquad  K(t)=K_0 t^{\gamma}
\end{equation}
The relations between the exponents are straightforward:
\begin{equation}
\alpha = \frac{\gamma + 1}{2 + \theta} \quad \mbox{  and } \quad \alpha + \beta = 1 + \epsilon
\end{equation}
Eq. (\ref{gen-eq}) can be understood by rescaling the time following $ \tau = t^{1+\gamma} $ in terms of the initial 
diffusion equation (\ref{diffeq}):
\begin{equation}
\frac{\partial \rho}{\partial \tau}= \frac{\partial}{\partial z} \left(\frac{K_0}{1+\gamma}\frac{1}{z^\theta} \frac{
\partial \rho}{\partial z}\right) 
\label{diffeqt2}
\end{equation}
And we obtain the stretched exponential behavior for large $z$ in agreement with the experimental measurements:
\begin{eqnarray}
\rho(z,t) \propto  t^{\beta} 
\exp \left[ -\frac{(1+\gamma)z^{2+\theta}}{K_0(2+\theta)^2 t^{1+\gamma}} \right]
\label{soleqdifft}
\end{eqnarray}
Within this general approach, we can now describe our experimental results. For simplicity's sake and to illustrate the 
model with no loss of generality, we estimate from the experimental fits that $ \theta=-\frac{3}{2}$, $\alpha=0.4$ and 
$\beta=0.4$ which leads to: $$ \gamma=-0.8  \qquad \mbox{  and } \qquad  \epsilon=-0.2$$
This corresponds to a time decreasing diffusion coefficient such as $t^{-0.8}$ and of the flux at $z=0$ such as $t^{-0.2}$.
The diffusion coefficient is expected to decrease versus time, while the porous medium is gradually filled by the 
platinum atoms. This leads to a more difficult diffusion. At the $z=0$ interface, 
the growing external layer provides an additional source for the diffusion of atoms inside the porous substrate. 
Moreover a decreasing flux indicates that the growing external layer is gradually screening the influence of the 
plasma, and that diffusion becomes less efficient due to pore filling as mentioned above. These effects are also 
observed at 5 Pa, with $ \theta=-\frac{4}{3}$, $\alpha=0.5$ and $\beta=0.25$ which leads to: $ \gamma=-0.83
\mbox{ and } \epsilon=-0.25$. These last two scaling exponents are very close to those at 0.5 Pa. Consequently, the self similar exponents depend on the deposition conditions whereas the critical exponents of the diffusion are unchanged. This means that the diffusion equations (\ref{gen-eq}-\ref{BCgen}-\ref{diffeqt2}) are robust 
enough to provide a general frame for deposition/diffusion mechanisms. 

On the basis of experimental results and a general diffusion model, we have proposed a new scenario for mass deposition 
and diffusion on a porous substrate. As the flux of sputtered atoms cannot be absorbed 
immediately by the substrate, an overlayer grows. The combined effects of the deposition that change the 
porosity properties of the substrate and of this growing layer that can act as a reservoir for diffusion can be 
accounted for by generalizing a standard model of diffusion in a porous medium with constant diffusive properties 
($\theta$ is constant). This allows time dependent 
coefficients to be introduced. This model provides a clear understanding of the experimental results: in 
particular, we show that the diffusion coefficient in the porous substrate and the flux at the interface decrease with 
time due to the atom deposition in the substrate. This general model can be also be applied to many other diffusion processes where the inner porous structure of a material is modified by a penetrating flux of matter, as for example, for filtration through porous membranes, sedimentation processes, ...

The CNRS is acknowledged for a Ph.D. scolarship (A. Caillard). Universit\'e d'Orl\'eans (BQR 2002) and  the R\'egion 
Centre is acknowledged for funding of the experimental setup. C.J. thanks the DGA for financial support. M. Mikikian is acknowledged for a careful reading of the manuscript. The reviewers are gratefully acknowledged for useful comments.

\end{document}